\documentclass[fleqn,usenatbib]{mnras}
\usepackage{graphicx}	
\usepackage{amsmath}	
\usepackage[dvipsnames]{xcolor}  
\usepackage{amssymb}	
\usepackage{newtxtext,newtxmath}
\usepackage{ulem}
\usepackage{macros_vdb} 
\defcitealias{VandenBosch2018a}{BO18}

\begin{document}
%


\title[Dynamical self-friction]
      {Dynamical self-friction: how mass loss slows you down}

\author[T. B. Miller et al.]{%
   Tim B. Miller,$^{1}$\thanks{E-mail: \href{tim.miller@yale.edu}{tim.miller@yale.edu} (TBM)}
   Frank~C.~van den Bosch,$^{1,2}$
   Sheridan~B.~Green,$^2$\thanks{NSF Graduate Research Fellow}
   and
   Go Ogiya$^{3,4,5}$
\vspace*{8pt}
\\
   $^1$Department of Astronomy, Yale University, PO. Box 208101, New Haven, CT 06520-8101\\
   $^2$Department of Physics, Yale University, PO. Box 208120, New Haven, CT 06520-8120\\
   $^3$Waterloo Centre for Astrophysics, University of Waterloo, Waterloo, ON N2L 3G1, Canada \\
    $^4$Department of Physics and Astronomy, University of Waterloo, 200 University Avenue West, Waterloo, Ontario N2L 3G1, Canada \\
   $^5$Laboratoire Lagrange, Universit\'e C\^ote d'Azur, Observatoire de la C\^ote d'Azur, CNRS,\\ \quad Boulevard de l'Observatoire, CS 34229, 06304 Nice, France
   }


\date{}

\pagerange{\pageref{firstpage}--\pageref{lastpage}}
\pubyear{2019}

\maketitle

\label{firstpage}


\begin{abstract}
We investigate dynamical self-friction, the process by which material that is stripped from a subhalo torques its remaining bound remnant, which causes it to lose orbital angular momentum. By running idealized simulations of a subhalo orbiting within an analytical host halo potential, we isolate the effect of self-friction from traditional dynamical friction due to the host halo. While at some points in a subhalo's orbit the torque of the stripped material can boost the orbital angular momentum of the remnant, the net effect over the long term is orbital decay regardless of the initial orbital parameters or subhalo mass. In order to quantify the strength of self-friction, we run a suite of simulations spanning typical host-to-subhalo mass ratios and orbital parameters. We find that the time-scale for self-friction, defined as the exponential decay time of the subhalo's orbital angular momentum, scales with mass ratio and orbital circularity similar to standard dynamical friction. The decay time due to self-friction is roughly an order of magnitude longer, suggesting that self-friction only contributes at the 10 percent level. However, along more radial orbits, self-friction can occasionally dominate over dynamical friction close to pericentric passage, where mass stripping is intense. This is also the epoch at which the self-friction torque undergoes large and rapid changes in both magnitude and direction, indicating that self-friction is an important process to consider when modeling pericentric passages of subhaloes and their associated satellite galaxies.
\end{abstract}




\section{Introduction}

Dynamical friction is an important astrophysical process. It causes dark matter subhaloes, and their associated satellite galaxies, to sink towards the centre of their host haloes and is ultimately responsible for the merging of galaxies and massive black holes in galaxy centres. In its classical treatment by \citet{Chandrasekhar1943}, which is formally only valid for a uniform, isotropic distribution of field particles, dynamical friction arises from the momentum and energy transfer from the heavy subject mass to the much less massive field particles during gravitational encounters. In a centrally concentrated mass distribution, such as a galaxy or a dark matter halo, dynamical friction arises from field particles that are in resonance with the subject mass. These resonant orbits exert a net retarding torque on the subject, causing it to sink towards the centre of the host system \citep[][Banik \& van den Bosch, in prep.]{Tremaine.Weinberg.84, Kaur.Sridhar.18}.

If the subject mass is an extended object (i.e., a dark matter subhalo or a satellite galaxy), it will also be subject to tidal forces from the host that can strip the subject of some of its mass. This stripped material typically is distributed in leading and trailing arms that are stretched out over time due to phase-mixing. As has been pointed out in a few studies \citep[e.g.,][]{Fujii2006, Fellhauer2007, VandenBosch2018a, Ogiya2019}, this stripped material can also exert a torque on the subject mass, thereby giving rise to a phenomenon that we henceforth refer to as (dynamical) self-friction.

Self-friction results in an enhanced dynamical friction force and thus in a shorter dynamical friction time, $\tau_{\rm df}$, defined as the time-scale on which the subject loses (some fraction of) its specific orbital angular momentum. This is automatically accounted for in studies that quantify the dynamical friction time of extended objects using numerical simulations \citep[see e.g.,][]{Velazquez1999, Jiang2000, Boylan-Kolchin2008, Jiang2008a}. However, it is not accounted for in analytical attempts to quantify how $\tau_{\rm df}$ scales with halo mass, orbital parameters, and other relevant properties (i.e., concentration of host and subhalo). The latter typically rely on Chandrasekhar's description for the dynamical friction force from the host halo, ignoring self-friction and using the local density of the host to estimate the instantaneous friction force.

These analytical estimates of $\tau_{\rm df}$ are used in semi-analytical models of dark matter substructure \citep[e.g.,][]{Taylor2001, Taffoni2003, Zentner2005} and in semi-analytical models of galaxy formation, where they are equated to the time-scale on which galaxies merge \citep[e.g.,][]{Kauffmann1999, Somerville1999, Cole2000} or they are used to estimate the merging time of `orphan' galaxies \citep[e.g.,][]{DeLucia2007, Kitzbichler2008}. They are also used in models and simulations of the merging of super-massive black holes \citep[SMBHs; e.g.,][]{Begelman1980, Fiacconi2013, Hirschmann2014, Tremmel2015}. In the case of SMBHs, self-friction is obviously not of concern, but whenever the subject mass is extended, the neglect of self-friction is likely to cause an overestimate of $\tau_{\rm df}$. However, since Chandrasekhar's treatment is not guaranteed to be accurate for a non-uniform density distribution such as a halo or galaxy anyway, the analytical estimates of the dynamical friction time are not expected to be particularly reliable. As a consequence, authors often introduce a multiplicative fudge factor that they tune to either obtain the correct morphological mix of galaxies\footnote{In semi-analytical models, it is typically assumed that mergers between galaxies of comparable mass results in the formation of an elliptical.} or to reproduce the dynamical friction times observed in numerical simulations. This fudge factor thus accounts for the potential impact of self-friction, making its neglect of little consequence.

However, if we ever want to move towards a more accurate analytical treatment of dynamical friction, it will be important to have some understanding of the relative contribution due to self-friction, as this would allow testing of the analytical models against numerical simulations. Furthermore, having a better understanding of self-friction is interesting in its own right, even if it is mainly academic. The first study to investigate self-friction, although they did not refer to it as such, was by \citet{Fujii2006}. Using an idealized simulation of a self-consistent $N$-body satellite system orbiting within a self-consistent $N$-body host halo, the authors showed that the stripped material indeed exerts a net torque on the bound remnant, which accounts for about 20 percent of the total drag force. In addition, the authors showed that the stripped material has an additional, indirect effect: since phase mixing is a slow process, material that was stripped from the subject mass relatively recently remains close to the remnant for quite some time. Hence, the perturbation that the field particles experience, which gives rise to a `wake' that is ultimately responsible for the retarding torque on the subject mass \citep[e.g.,][]{Mulder1983, Kalnajs1971, Colpi1999}, is proportional to the subject's bound mass plus some of its (recently) stripped mass. This was confirmed by \citet{Fellhauer2007}, who suggested that, as a rule-of-thumb, the magnitude of this indirect effect can be approximated as coming from half of the mass that has become unbound during the preceding orbit. 

The simulations used by \citet{Fujii2006} followed the orbital evolution of a self-consistent satellite galaxy (without dark matter). As they point out in their paper, since dark matter subhaloes are more extended than satellite galaxies, they typically experience more mass loss and thus should be susceptible to even stronger self-friction. In this paper we use idealized numerical experiments to study the nature of self-friction acting on dark matter substructure. Our main goal is to characterize the relative contribution of self-friction to the total drag force experienced by subhaloes as a function of orbital properties and the mass ratio between the sub- and host halo. 

The rest of the paper is organized as follows: in Section~\ref{sec:meth}, we discuss our numerical methods and simulations. Our results are presented in Section~\ref{sec:res} and then discussed and summarized in Section~\ref{sec:disc}. Throughout,  we  adopt  a  Hubble  parameter $H_0 = 70\ \rm km\, s^{-1}\, Mpc^{-1}$, which corresponds to a Hubble time of $13.97$ Gyr for a flat universe with $\Omega_{\Lambda} = 0.73$.

\begin{figure*}
\centering
\includegraphics[width = 0.99\textwidth]{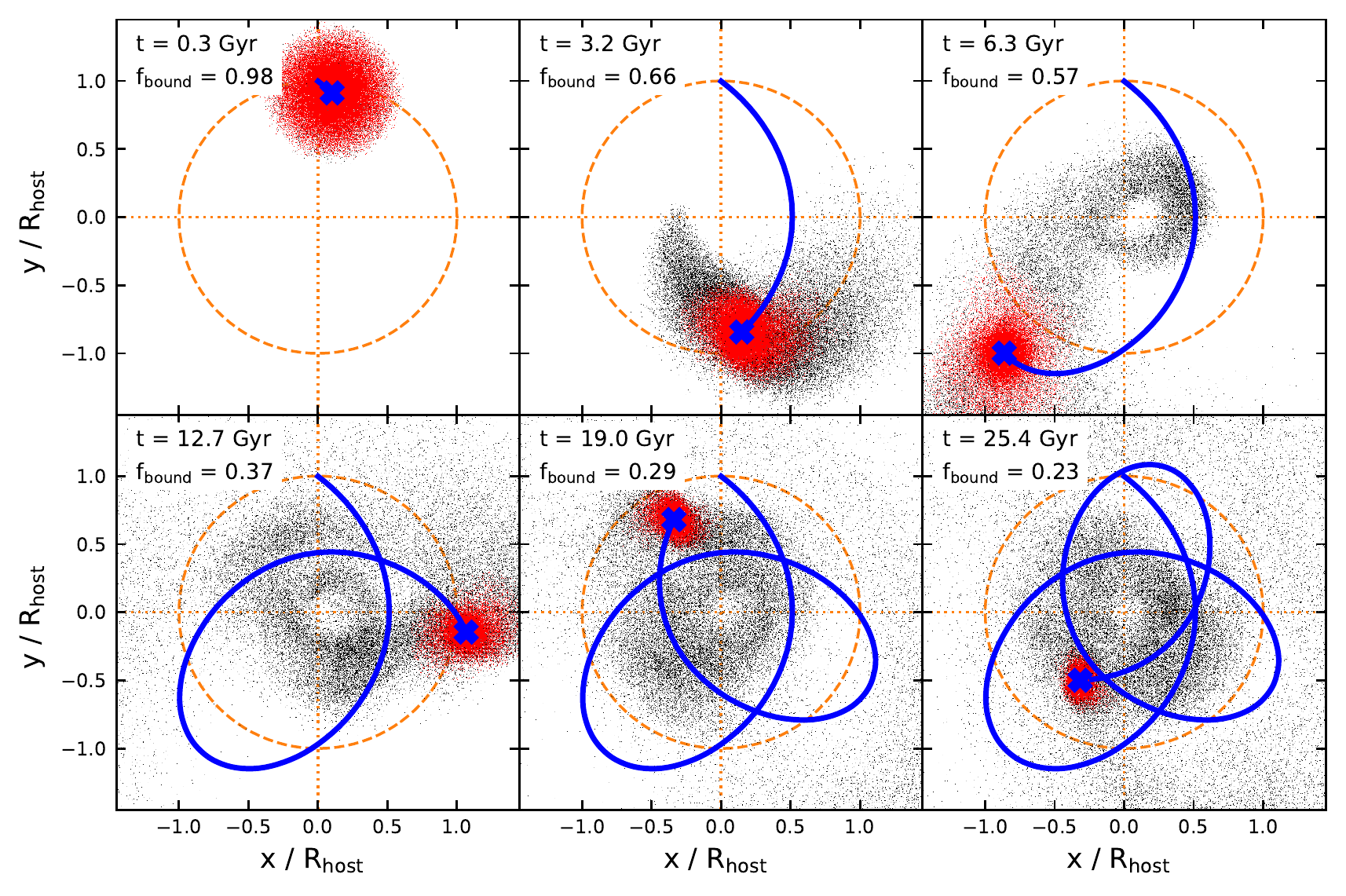}
\caption{The bound (red) and unbound (black) particles of a subhalo at six snapshots throughout its evolution. The subhalo is initialized on an orbit with $\eta = 0.8$, $x_\rmc = 1$, and $\logmhs = 1$. This subhalo is initialized with $r_{\rm trunc} =  r_{\rm vir,sub}$. The blue cross displays the current COM position of the subhalo and the blue line displays the orbital trajectory leading up to each snapshot. The time since the start of the orbit and the subhalo's bound mass fraction, $f_\rmb$, are indicated in each panel. The orange dashed circle indicates the virial radius, $R_{\rm host}$, of the host halo, while the orange dotted lines indicate its center.}
\label{fig:intro}
\end{figure*}

\section{Methods}
\label{sec:meth}

To investigate the effects of self-friction, we perform numerical simulations of an $N$-body subhalo orbiting within an analytical host halo potential. We provide a brief description of the methods here and refer the reader to \citet[][hereafter \citetalias{VandenBosch2018a}]{VandenBosch2018a}  for full details on the numerical simulations and analysis methods.

The host- and subhaloes are both assumed to initially be spherical NFW profiles \citep{Navarro1996} with concentrations $c_{\rm host}$ and $c_{\rm sub}$, respectively. The host halo is represented by a static, analytical potential. Throughout this paper, $R$ and $r$ are used to indicate radii with respect to the center of the host halo and subhalo, respectively. The virial masses and radii of the two systems\footnote{Throughout this paper, the virial radius of a halo is defined as the radius that encloses an average density equal to 97 times the critical density for closure at the present day.} are indicated by $M_{\rm host}$ and $R_{\rm host}$, in the case of the host halo, and $M_{\rm sub}$ and $r_{\rm sub}$ in the case of the subhalo. Note that for the latter, these relate to the initial, unperturbed subhalo, prior to being introduced to the tidal field of the host halo. 

Positions and velocities of particles in the subhalo are initialized assuming an isotropic distribution function, $f(E)$ using the method of \citet{Widrow2000}. In order to roughly account for the tidal mass loss that the subhalo would have experienced when approaching its initial position from an infinite distance, the initial distribution of particles is truncated at the instantaneous tidal radius of the subhalo \citep{King1962}, given by 
\begin{equation}
r_{\rm tidal} = R \left( \frac{m(r_{\rm tidal})/M(R) }{2 + \frac{\Omega^2 R^3}{G\,M(R)} - \frac{\rmd \ln M}{\rmd \ln R}\big|_R } \right)^{1/3}.
\end{equation}
Here, $R$ is the (initial) distance from the center of the subhalo to the center of the host halo, $m(r_{\rm tidal})$ is the mass of the subhalo within $r_{\rm tidal}$, $M(R)$ is the mass of the host halo within $R$, and $\Omega$ is the (initial) angular speed of the subhalo~\citep{Taylor2001, Zentner2003, Penarrubia2005, VandenBosch2017}. Note that we do not account for this truncation in the distribution function, which instead is computed under the assumption that the subhalo's density profile extends to an infinite radius. This implies that the initial subhalo is not perfectly in equilibrium. We emphasize that due to the tidal field within which it is embedded, the subhalo is not expected to be in perfect equilibrium anyway. Furthermore, \citetalias{VandenBosch2018a} have demonstrated that truncating the initial subhalo at the tidal radius or at its formal virial radius only has a very small impact on the subsequent mass evolution of the subhalo. As we demonstrate in \S\ref{sec:case} below, this choice only impacts self-friction, the topic of this paper, at a few percent level. 

The initial orbit of a subhalo is characterized by two parameters, $x_\rmc$ and $\eta$. These quantities are related to the canonical measures of orbital energy, $E$, and specific angular momentum, $j$, according to
\begin{equation}
x_c = \frac{R_{\rm circ}(E)}{R_{\rm host}}\,\,\,\,\,\,\,\,\,\text{and}\,\,\,\,\,\,\,\,
\eta =  \frac{j}{j_{\rm circ}(E)}\,.
\end{equation}
Here, $R_{\rm circ}(E)$ is the radius of a circular orbit with energy $E$, and $j_{\rm circ}(E)$ is the specific angular momentum of a circular orbit with energy $E$. The parameter $\eta$ is known as the orbital circularity. Each simulation is uniquely specified by $c_{\rm host}$, $c_{\rm sub}$, $x_\rmc$, $\eta$, and the initial ratio of the virial masses of the host halo and subhalo, $M_{\rm host}/M_{\rm sub}$. The host and subhalo concentration are fixed in our simulations at values of $c_{\rm host}=5$ and $c_{\rm sub}=10$. Unless otherwise specified, the subhalo is initialized at the virial radius of the host halo while approaching pericenter.

All simulations are performed using a modified version of the hierarchical $N$-body code \textsc{treecode} written by Joshua Barnes with some improvements made by John Dubinski. It uses a \citet{Barnes1986} octree to compute accelerations based on a multipole expansion up to quadrupole order and uses a second-order leap-frog integration scheme to solve the equations of motion. Forces between particles are softened using a simple Plummer softening. Throughout this paper, we work in simulation units where the gravitational constant, $G$, the initial scale radius of the subhalo, $r_{\rm s,sub} \equiv r_{\rm sub}/c_{\rm sub}$, and the initial virial mass of the subhalo, $M_{\rm sub}$, are all unity. Unless otherwise specified, simulations are ran until $t = 31.6$ Gyr. The fiducial numerical parameters are summarized in Table~\ref{tab:num}. As we show in Appendix~\ref{app:num}, these parameters are adequate to properly resolve the self-friction that is central to this study. 

The output of every 100$^\textrm{th}$ timestep  (every 0.127 Gyr) is saved as a snapshot and used for analysis, leading to 250 snapshots for each simulation. At each snapshot, we evaluate whether a particle is still bound to the subhalo according to a slight modification of the procedure laid out in \cite{VandenBosch2018} and \citetalias{VandenBosch2018a}. This algorithm iteratively identifies particles as bound or unbound (i.e., if its binding energy to the subhalo is negative or positive), recalculating the binding energy until the center-of-mass (COM) position, $\bR_{\rm com}$, and velocity, $\bV_{\rm com}$, of the subhalo are stable to within 0.5\% of the initial virial radius and velocity, respectively. Additionally, we do not allow particles to ``re-bind'' to the subhalo if they have previously been unbound for longer than $0.15$ Gyr (i.e., roughly the time between two snapshots).
This prevents particles that are far away from the subhalo, but that happen to be moving at roughly the same speed as the subhalo, to be considered bound. As noted in \citetalias{VandenBosch2018a}, this can occasionally occur for a small fraction of particles \citep[also see][]{Penarrubia2008}. And although this does not significantly affect the bound mass fraction, it can have a non-negligble effect on the measurement of $\bR_{\rm com}$ and $\bV_{\rm com}$. In particular, we find that it can cause `fluctuations' in the time evolution of the subhalo's orbital angular momentum that can be as large as $10-15\%$. Not allowing for the re-binding solves this issue.

At each simulation output, we use the bound particles thus defined to compute both the subhalo's specific, orbital angular momentum $\bj = \bR_{\rm com} \times \bV_{\rm com}$, as well as its bound fraction $f_\rmb \equiv M_{\rm sub}(t)/ M_{\rm sub}$. Here, $M_{\rm sub}(t)$ is the total mass of bound subhalo particles at time, $t$. Note that the initial bound fraction at $t=0$ is less than unity since the initial subhalo is truncated at its instantaneous tidal radius, rather than its virial radius. Typically, the initial bound fraction is $\sim 0.9$, with a slight dependence on the initial orbital parameters. Simulations are analyzed until the bound mass fraction falls below $2 \times 10^{-2}$. This limit is set by the numerical reliability guidelines laid out in \citetalias{VandenBosch2018a}, as appropriate for our fiducial numerical parameters (see Table~\ref{tab:num}). Below this bound fraction, the results become numerically unreliable due to the effects of discreteness noise and inadequate force softening. 
\begin{table}
	\centering
	\begin{tabular}{lll} 
		\hline
		Parameter & Symbol & Value\\
		\hline
		Number of particles & $N_\rmp$ & $10^5$\\
		Opening angle & $\theta$ & $0.7$\\
		Time step  & $\Delta t$ & 0.02\\
		Softening length & $\epsilon$ & 0.05\\
        \hline
	\end{tabular}
	\caption{The fiducial numerical parameters of the simulations. The time step and softening length are in simulation units ($G = M_{\rm sub} = r_{\rm s, sub} = 1$).}
	\label{tab:num}
\end{table}

\section{Results}
\label{sec:res}

\subsection{A case study}
\label{sec:case}

To develop some insight, we start by analyzing one specific simulation in detail. Fig.~\ref{fig:intro} shows the distribution of particles at six snapshots during the evolution of a subhalo initialized with $\logmhs = 1$, $x_\rmc = 1$ and $\eta = 0.8$. At each snapshot, the bound and unbound particles are indicated by red and black dots, respectively. The blue curve indicates the past orbital trajectory and the blue cross marks the COM of the bound subhalo remnant at the time of the snapshot. During its first pericentric passage (at $t = 1.8$ Gyr), the subhalo loses about $15\%$ of its mass. The amount of mass lost increases to $\sim 70\%$ after one Hubble time. The tidally stripped material from the leading and trailing arms is phase-mixed into a long stream which wraps in on itself, creating a roughly donut-shaped structure. 
\begin{figure}
\centering
\includegraphics[width = \columnwidth]{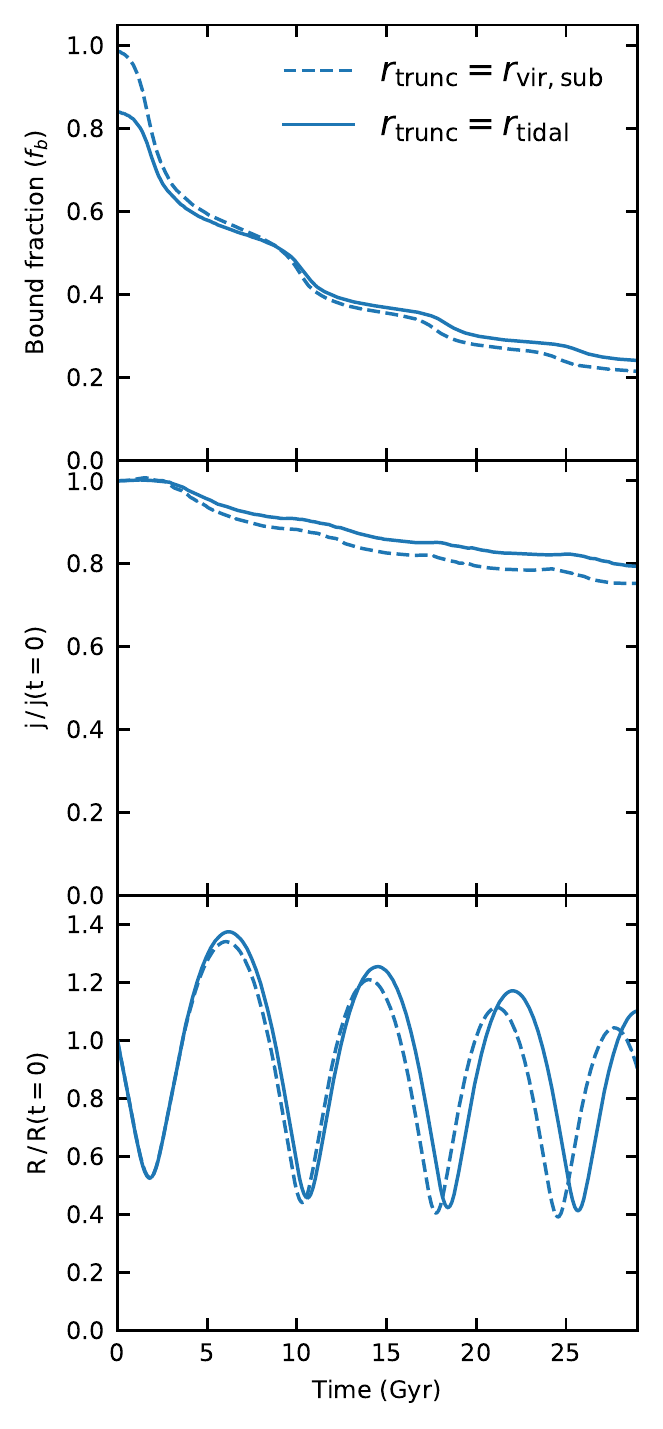}
\caption{The evolution of the bound mass fraction $f_\rmb$ (top panel), specific orbital angular momentum $j$ (middle panel), and the orbital radius $R$ (bottome panel) for the subhalo shown in  Fig.~\ref{fig:intro}. The dashed lines show the results for a subhalo with the same virial mass and on the same orbit, but whose initial extent is truncated at the virial radius, rather than the tidal radius. This results in more mass loss during the first pericentric passage, which subsequently causes a somewhat stronger self-friction and thus a more pronounced loss of orbital angular momentum. }
\label{fig:intro_orbit}
\end{figure}

While not readily apparent in Fig.~\ref{fig:intro}, the specific orbital angular momentum, $j = |\bj|$, of the subhalo declines with time. Fig.~\ref{fig:intro_orbit} shows this more clearly. The solid lines show the time evolution of the bound mass fraction $f_\rmb$ (upper panel), the specific orbital angular momentum $j$ (middle panel), and the orbital radius $R$ (lower panel). Both $j$ and $R$ are normalized to their initial values. Note that the subhalo experiences tidal mass loss as well as a loss of specific orbital angular momentum, causing a net decay of orbital radius with time. Recall that these simulations are run with a static, analytical potential representing the host halo. Hence, the orbital decay is not due to dynamical friction caused by constituent particles of the host halo. Rather, the orbital decay is a manifestation of self-friction \textemdash the gravitational back-reaction of stripped material on the bound remnant of its own parent subhalo. In the case of the simulation shown in Fig~\ref{fig:intro} and~\ref{fig:intro_orbit}, this self-friction has `robbed' the subhalo of about 13 (20) percent of its initial, specific angular momentum after 10 (20) Gyr.

The dashed lines in Fig.~\ref{fig:intro_orbit} show the results obtained using a simulation with the same orbital parameters and the same value of $\mhs$, but in which the initial subhalo is truncated at its virial radius, $r_{\rm sub}$, rather than its instantaneous tidal radius, $r_{\rm tidal}$. Initially, the bound mass fraction of this subhalo is unity and thus larger than in the fiducial case. This extra mass is rapidly stripped off during the first pericentric encounter and after 10 Gyr the bound mass is similar to that of the fiducial simulation. However, since the subhalo has now lost more mass, it also experiences enhanced self-friction (i.e., a more rapid decline of the specific orbital angular momentum). Consequently, the subhalo comes closer to the center of the host halo, experiencing a stronger tidal force. This in turn boosts the mass loss rate and, as a consequence, the bound fraction of the subhalo at late times is somewhat {\it lower} than in the fiducial case. We have performed similar tests for different orbital parameters and different mass ratios, always finding that truncation at the virial radius results in a somewhat more rapid decline of $j(t)$. However, the differences never exceed $5\%$ within the time span of our simulations ($\sim$ 30 Gyr). Since the main goal of this paper is to develop insight into the effect and scaling of self-friction, rather than construct an accurate model to predict its outcome, the details about how the initial subhalo is truncated are inconsequential for what follows.

Finally, as we demonstrate in Appendix~\ref{app:num}, these results are robust to changes in the softening length and are not affected by discreteness noise. The results are converged, in the sense that running the simulations with more particles yields indistinguishable results. 

\subsection{Self-friction demographics}
\label{sec:demo}

\begin{figure*}
\centering
\includegraphics[width = \textwidth]{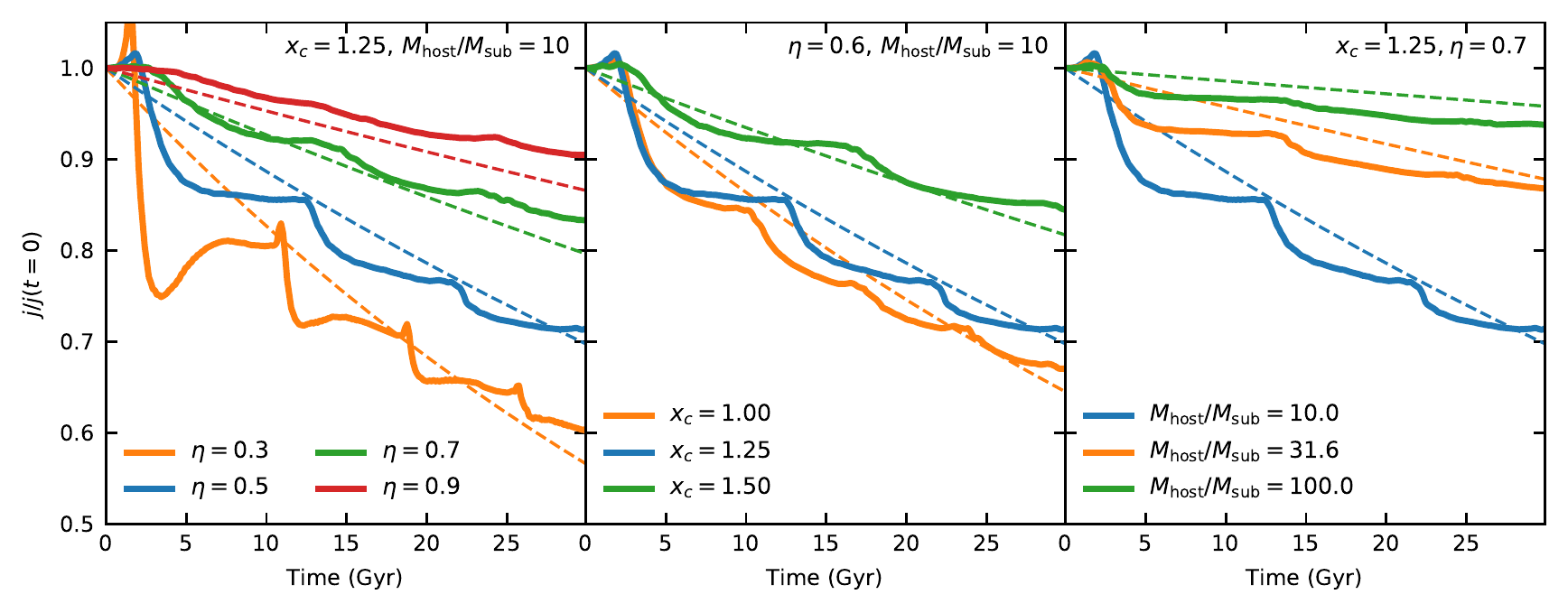}
\caption{The evolution of the specific angular momentum for three sets of orbits. In each set, we vary one of $\eta$ (left), $x_\rmc$ (center) or $\mhs$ (right)  and keep the other two fixed. Shown in the dotted line is the exponential decay model, described by equation~\eqref{eqn:j_fit} and Table~\ref{tab:fit}, for each specified orbit. The exponential decay model does not capture the detailed evolution of each individual simulation, but the overall rate of decay as a function of time and orbital parameters is well-approximated. The temporal increases of $j$ during the first 1-2 Gyr noticeble for the more radial orbits is likely to be an artefact due to the fact that the potential of the host halo is kept fixed; if the host halo were live, its centre would move with respect to the common centre of mass \citep[see e.g.,][]{Ogiya2016}.}
\label{fig:j_fit}
\end{figure*}

The solid lines in Fig.~\ref{fig:j_fit} show how the evolution of the specific angular momentum due to self-friction depends on the orbital circularity $\eta$, the initial mass rate $\mhs$, and the orbital energy as characterized by $x_\rmc$. In each case, we vary these parameters with respect to a fiducial case, which has $x_\rmc = 1.25$, $\eta=0.5$, and $\mhs=10$ (indicated by the blue curve in each panel).  Note that self-friction is more pronounced for more eccentric (lower $\eta$) orbits, for orbits that are more bound (lower $x_\rmc$), and when the mass ratio $\mhs$ is smaller, i.e. a more massive subhalo. Note also that the evolution of $j$ can be very erratic at times, especially for the more eccentric orbits. 

\begin{figure*}
\centering
\includegraphics[width = \textwidth]{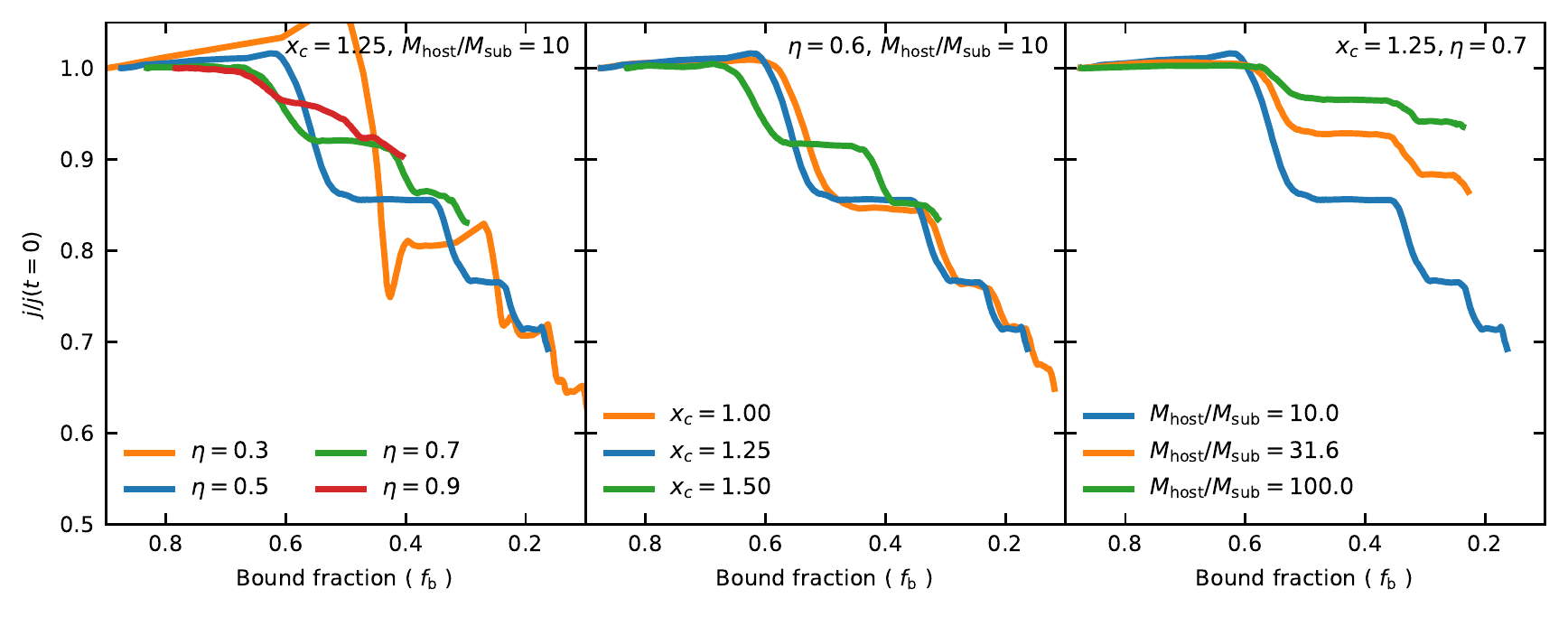}
\caption{The evolution of the specific angular momentum $j$  as a function of bound fraction. Note that the bound fraction decreases to the right on the horizontal axis so that it can be interpreted similarly to Fig.~\ref{fig:j_fit}. Interestingly, we find that changing the orbital parameters, $x_\rmc$ and $\eta$, does not affect the evolution in $j$ - $f_\rmb$ space. All of the orbits in the first two panels seem to lie on a similar track, suggesting that the main determinant of self-friction is how much material has been stripped. However, this is not the case in the rightmost panel, when orbits of different initial mass ratios are compared. As we discuss in \S~\ref{sec:demo}, for a more massive subhalos the force from the stripped particles represents a larger fraction of the total force on the subhalo }
\label{fig:fb_j}
\end{figure*}

As we demonstrate explicitly below, self-friction arises from the torque exerted on the bound remnant by its stripped material. Placing a subhalo on a more-bound or more-eccentric orbit implies a smaller pericentric radius and thus more mass loss. This in turn implies a stronger torque and therefore, more self-friction. Fig.~\ref{fig:fb_j} shows the evolution of the specific angular momentum for the same simulations as in Fig.~\ref{fig:j_fit}, except that now the $j$ evolution is plotted as a function of the instantaneous bound mass fraction, $f_\rmb$, rather than time. The curves in the left and middle panels, which show the dependencies on $\eta$ and $x_\rmc$, respectively, are now much more similar. This indicates that the dependence of self-friction on orbital parameters mainly comes from a dependence on $f_\rmb$, which in turn has a strong dependence on $\eta$ and $x_\rmc$. The evolution of $j(t)$ as function of $f_\rmb$ is not entirely self-similar, though, with strong deviations being evident, especially during the first orbital period. As we discuss in \S~\ref{sec:phys}, these deviations arise from a complicated time-dependence of the angle between $\bj(t)$ and the torque vector. As is evident from the right-hand panel of Fig.~\ref{fig:fb_j}, the evolution of $j(t)$ as function of $f_\rmb$ is not self-similar for different mass ratios, $\mhs$, to the extent that in host halos of a given mass, more massive subhaloes lose more of their specific orbital angular momentum due to self-friction. This is easy to understand; for more massive subhaloes the force from its stripped material constitutes a larger fraction of the total force (including that of the host).

To conclude, for a given host halo, more massive subhaloes are more strongly affected by self-friction and at any point in time, the amount of specific angular momentum that a subhalo has lost due to self-friction is tightly correlated with its instantaneous bound mass fraction. Put differently, the rate at which self-friction robs the subhalo of its specific angular momentum is directly proportional to its mass loss rate, with a constant of proportionality that increases with decreasing $\mhs$. 

\subsection{Self-friction merging timescale}
\label{sec:ts}

In order to more quantitatively assess the impact of self-friction, we investigate how the `merging timescale' for self-friction depends on the initial parameters of the subhalo. Following previous works, we define the merging timescale based on the specific angular momentum, $j$, rather than radius, as it gives more reliable results, especially for highly eccentric orbits \citep{Boylan-Kolchin2008}. Often the merging timescale is defined to be the time when the angular momentum of the satellite reaches zero; however, it is computationally infeasible for us to run simulations up to this point. Rather, we define the self-friction timescale as the exponential decay time, $\tau_{\rm SF}$, of the specific orbital angular momentum, according to
\begin{equation}
\label{eqn:j_fit}
j(t) = j_0 \, e^{-t/\tau_{\rm SF}}\,.
\end{equation}
Here $j_0 = j(t=0) = \eta \, j_{\rm circ}(E)$ is the initial specific angular momentum of the subhalo's orbit of orbital energy $E$ and circularity $\eta$. Rather than fitting $\tau_{\rm SF}$ for each individual simulation, we fit the $j(t)$ profiles of all our simulations simultaneously, whereby we follow  \citet{Boylan-Kolchin2008} in adopting a $\tau$-dependence on orbital parameters and initial mass ratio given by
\begin{equation}
\tau_{\rm SF} = \tau_0 \, x_\rmc^{a} \, \exp(b \eta) \, \frac{(M_{\rm host} / M_{\rm sub})^{c}}{\ln ( 1 + M_{\rm host} / M_{\rm sub} )}\,.
\label{eqn:tau_fit}
\end{equation}
Here, $\tau_0$, $a$, $b$, and $c$ are free parameters whose best-fit values we infer by fitting the $j(t)$ profiles, as described below. Although this method relies on an assumed, exponential decay of the specific, orbital angular momentum, it has the advantage that it avoids the subjective choice of when a subhalo has `merged'. However, it implies that caution is required when comparing these $e$-folding times to the more standard dynamical friction times used in most other studies \citep[e.g.,][]{Lacey1993,Boylan-Kolchin2008,Jiang2008a}.

\begin{figure}
	\centering
    \includegraphics[width = \columnwidth]{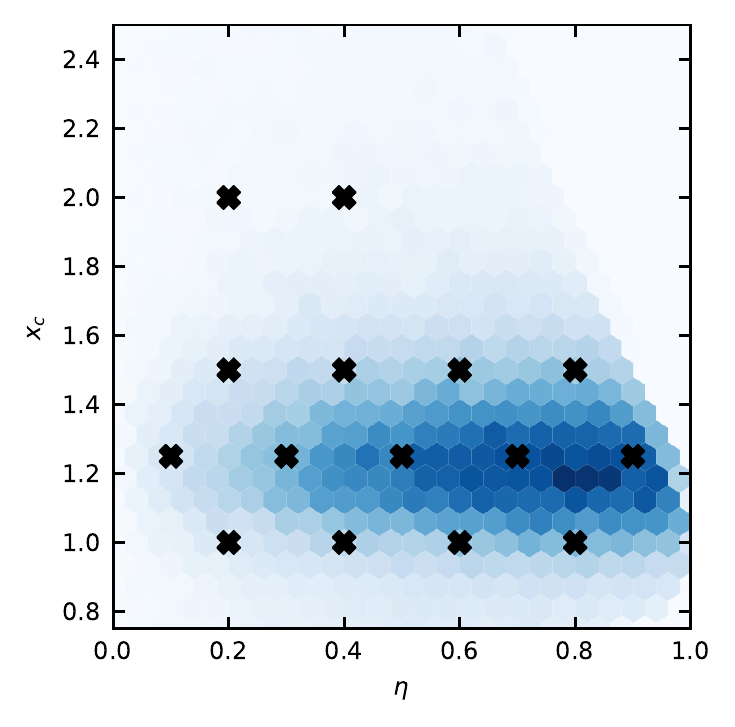}
    \caption{The shaded hexagons indicate the distribution of $x_\rmc$ and $\eta$ of infalling satellites for a host halo of $10^{13} \Msun$ taken from \citet{Jiang2015}. Darker shading indicates a larger contribution of orbits. The black crosses show the set of initial orbital configurations used in our suite of simulations. For each configuration we run three simulations with $\logmhs = 1$, $1.5$, and $2$. When we perform our fit for the merging timescale of self-friction, we weight each orbital configuration based on the relative probability from the distribution of \citet{Jiang2015} shown in blue.}
    \label{fig:eta_xc_dist}
\end{figure}

To explore the parameter space of realistic subhaloes, we run a suite of simulations with orbital parameters that sample the distribution of $\eta$ and $x_\rmc$ of subhaloes at infall in cosmological $N$-body simulations. This distribution, taken from \cite{Jiang2015}, is indicated with shaded polygons in Fig.~\ref{fig:eta_xc_dist}. Although this particular distribution is formally only valid for a host halo of mass $10^{13} M_\odot$ and mass ratios of $20 < M_\mathrm{host} / M_\mathrm{sub,i} < 200$, the dependence on host halo mass and mass ratio is relatively weak \citep{Wetzel2011} and will be ignored in what follows. The black crosses indicate the combinations of $\eta$ and $x_\rmc$ for each of which we have run three idealized simulations with $\logmhs = 1$, $1.5$, and $2$. For each simulation we measure the evolution of the subhalo's specific angular momentum, $j(t)$. We then fit all these $j(t)$ simultaneously, for all of the simulations, using the parametrization given by equations~(\ref{eqn:j_fit}) and~(\ref{eqn:tau_fit}), and adopting the cost function
\begin{equation}
    \calC = \sum_{k = 0}^{N_k} \ \sum_{j = 0}^{N_j} \  \left[ \frac{j_k (t = t_j)}{ j_k (t = 0) } - e^{-t_j / \tau_{{\rm SF},k} } \right]^2 .
\end{equation}
Here, we sum over all the $N_k=45$ simulations and all $N_j = 250$ snapshots within each simulation, comparing the angular momentum measured from the simulation, $j_k (t = t_i)$, to the predictions from the exponential decay model. $\tau_{{\rm SF},k}$ is the exponential self-friction decay rate for the $k$-th simulation, which depends on the initial orbital parameters, host-to-subhalo mass ratio, and the free parameters $a$, $b$, $c$, and $\tau_0$. The cost function, $\calC$, is minimized to find the best fit free parameters using the Levenberg-Marquardt algorithm \citep{More1978}. In order to estimate uncertainties in the parameters, and to account for the non-uniform distribution of orbits in $(\eta,x_\rmc)$ space, we perform a bootstrap analysis, in which we randomly sample, with replacement, $N_k=45$ simulations from our complete set, adopting a sampling probability proportional to the relative probabilities from the $(\eta,x_\rmc)$ distribution of \cite{Jiang2015} shown as blue, shaded hexagons in Fig.~\ref{fig:eta_xc_dist}. For each realization, $\calC$ is minimized and the best fit values of $a$, $b$, $c$, and $\tau_0$ are recorded. The median and half of the $16\% - 84\%$ interval of $10^4$ bootstrap samples are reported as the best fit values and error bars in Table~\ref{tab:fit}. This method effectively weights each simulation by the expected frequency of its corresponding orbital parameters.

\begin{table}
	\centering
	\large{
    \begin{tabular}{l|l}
    Parameter & Best fit value\\
    \hline
	$\tau_0$ [Gyr] &  $2.55\pm 0.55$ \\
    $a$ & $1.91 \pm 0.27$ \\
    $b$ &  $2.29 \pm 0.24$ \\
    $c$ & $1.21 \pm 0.1$ \\
    \hline
	\end{tabular}
	}
	\caption{The best fit values for our model of $\tau$ (equation~[\ref{eqn:tau_fit}]), the $e$-folding time of the subhalo orbital angular momentum, $j$.}
    \label{tab:fit}
\end{table}

The behaviour of the self-friction merging timescale largely mirrors that of subhalo merging timescale in the literature. The roughly linear dependence on $\mhs$ is consistent with previous studies and the exponential scaling of $\eta$ is roughly the same as that of the timescale in \citet{Boylan-Kolchin2008} ($2.3\pm 0.2$ in this work as opposed to their $1.9$). The major difference is the dependence on orbital energy, which is characterized by the power law slope of $x_\rmc$. \Citet{Boylan-Kolchin2008} and \citet{Jiang2008a} report scalings of $\tau \propto x_\rmc^{1}$ and $\tau \propto x_c^{0.5}$, respectively, whereas we find a much steeper dependence of $x_\rmc^{1.91}$ for self-friction

The dashed lines in Fig.~\ref{fig:j_fit} show the predictions based on our exponential decay model using these best fit values. Overall, this simple model describes the evolution of the specific angular momentum fairly well. In particular, it nicely captures the overall trends with $\eta$, $x_\rmc$ and $\mhs$, and over the entire range of parameters shown. The model fails to account for the various sharp features in the individual $j(t)$ curves, something that becomes more acute for the most eccentric orbits. In principle, this could potentially be accounted for in a more sophisticated model in which the decay of angular momentum is a function of orbital phase, but this is beyond the scope of this paper. Overall, the simple exponential decay model adequately describes the qualitative evolution of $j$ in our simulations of self-friction.
\begin{figure*}
\centering
\includegraphics[width = \textwidth]{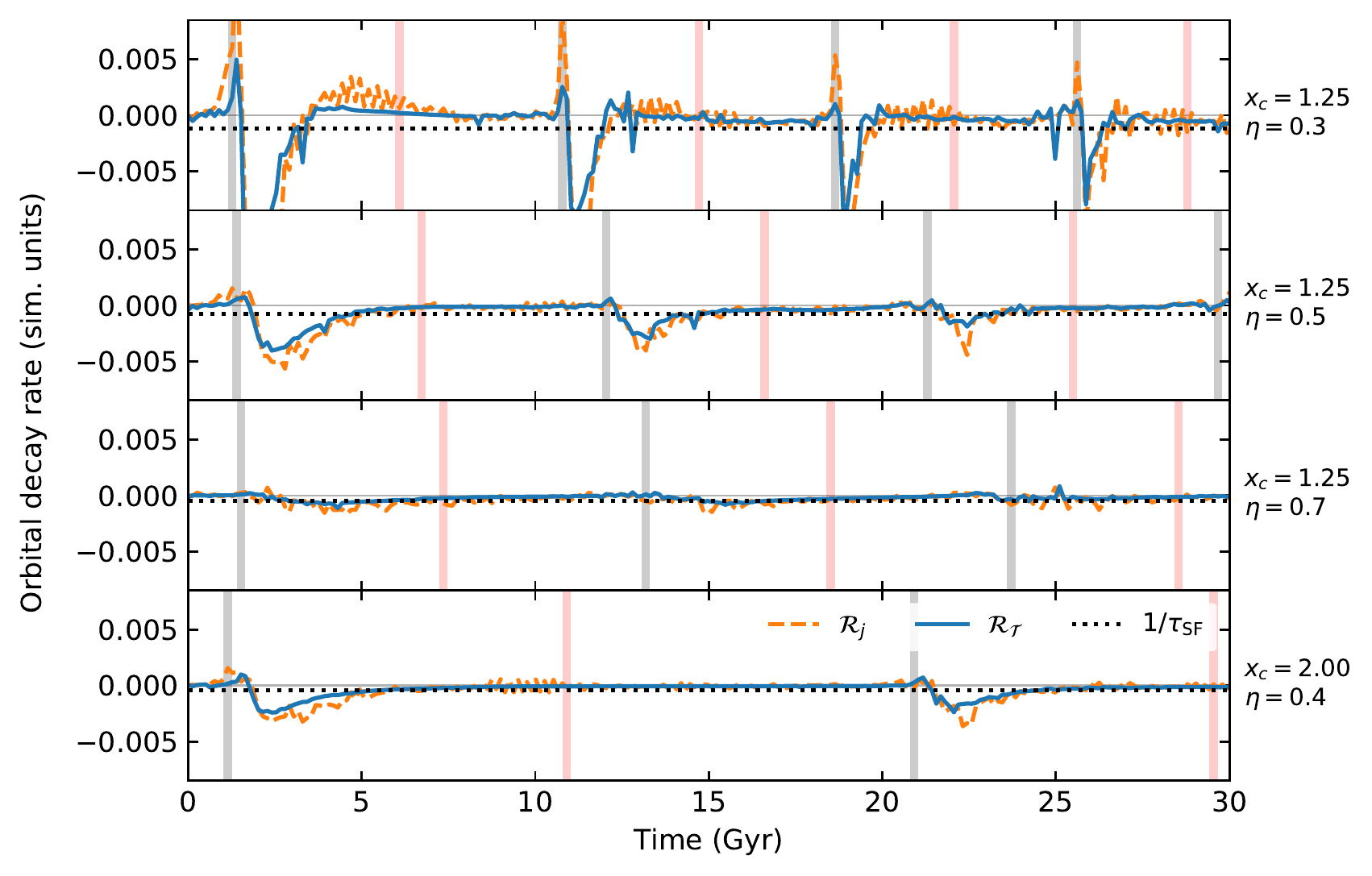}
\caption{The evolution of three different orbital decay rates for four different orbital configurations are shown. All four orbits are initialized with $\logmhs = 1$ and the different orbital parameters are shown in each panel. The orange dotted line shows the decay rate $\mathcal{R}_j$, which is calculated directly from the evolution of $j$ using a simple numerical derivative. The blue solid line shows the decay rate $\mathcal{R}_{\mathcal{T}}$ calculated using the torque caused by the previously stripped particles acting on the COM of the subhalo. The black dotted line shows a third rate calculated as $1 / \tau_{\rm SF}$. Here, $\tau_{\rm SF}$ is calculated using our best fit model described in Table~\ref{tab:fit} and equation~\ref{eqn:tau_fit} which is based on the initial conditions of each orbit. By definition, $1 / \tau_{\rm SF}$ is constant throughout the orbit. A thin grey line is shown at the value of zero in order to guide the eye. The grey and pink vertical lines show the times of pericentric and apocentric passages, respectively. We find that $\mathcal{R}_j$   matches $\mathcal{R}_{\mathcal{T}}$ over the entire 30 Gyr simulation for all four orbits. This further corroborates our hypothesis that it is the torque due to the stripped material acting on the subhalo that is responsible for the observed orbital decay.}
\label{fig:torque}
\end{figure*}

\subsection{The origin of self-friction}
\label{sec:phys}

The orbital decay of subhaloes arises from the action of a torque. In the case of self-friction, this torque is due to the subhalo material that has been stripped previously. We can get some insight by computing this torque and comparing it to the rate at which orbital angular momentum is being lost. The orange, dashed lines in Fig.~\ref{fig:torque} depict the rate at which subhaloes on different orbits (different panels) lose their specific, orbital angular momentum, as quantified by $\calR_j \equiv (1/j) \rmd j/\rmd t = \rmd\ln j/\rmd t$. Here $j(t)$ is measured directly from the simulations as described in \S\ref{sec:meth} above, and the derivative is calculated using a simple first-order finite difference scheme. The black dotted line corresponds to $1/\tau_{\rm SF}$, with $\tau_{\rm SF}$ calculated using equation~\eqref{eqn:tau_fit} and our best fit values from Table~\ref{tab:fit}. By definition, this is a constant throughout the orbit. In most cases, this provides a good description of the {\it average} rate of specific angular momentum loss. However, the actual orbital angular momentum loss rate, $\calR_j$, reveals pronounced fluctuations with time that are not captured by this average. In fact, these fluctuations can be positive, corresponding to an angular momentum gain! The fluctuations are most pronounced during pericentric passages, indicated by grey, vertical bars (pink vertical bars indicate apocentric passages). In particular, right around pericentric passage, the subhalo typically gains some specific orbital angular momentum, only to experience a drastic loss shortly thereafter. In general, the magnitude of these fluctuations are more pronounced for more eccentric orbits.

The solid blue curves in Fig.~\ref{fig:torque} indicate the expected rate, $\calR_{\calT}$, computed from the specific torque on the subhalo due to its own stripped material, $\bmath{\calT}_{\rm SF} = \bmath{R}_{\rm COM} \times \bmath{a}_{\rm stripped}$. Here  $\bmath{R}_{\rm COM}$ is the COM position of the subhalo from the center of the host, and $\bmath{a}_{\rm stripped}$ is the acceleration at this COM due to all the stripped (unbound) material. The rate is computed from the component of the torque in the direction of the instantaneous orbital angular momentum vector $\bj$ using
\begin{equation}
\calR_{\calT} = \frac{\bmath{\calT}_{\rm SF}}{j} \cdot \frac{\bj}{j} = \frac{\bmath{R}_{\rm COM} \times \bmath{a}_{\rm stripped}}{\vert \bmath{R}_{\rm COM} \times \bmath{V}_{\rm COM}\vert} \cdot \be_{j}
\end{equation}
with $\be_j$ the unit vector in the direction of $\bj$. As is evident, the actual rate, $\calR_j$, is very similar to the predicted rate, $\calR_{\calT}$, indicating that self-friction indeed arises from the torque caused by the stripped material. In particular, $\calR_{\calT}(t)$ nicely reproduces the fluctuations in the angular momentum loss rates, including the temporary boost in orbital angular momentum (corresponding to positive values of $\rmd\ln j/\rmd t$) right around pericentric passage.

\begin{figure*}
\centering
\includegraphics[width = \textwidth]{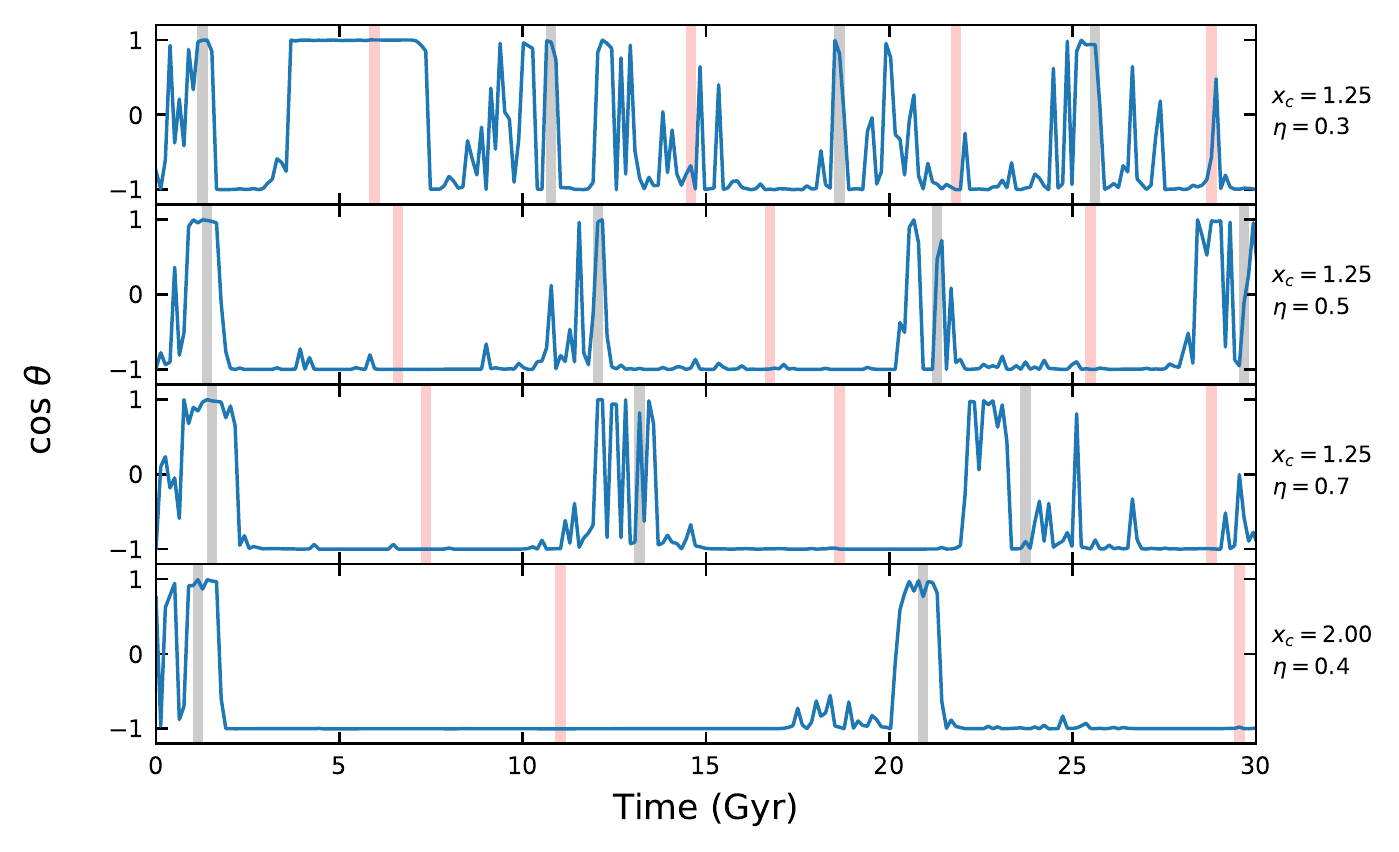}
\caption{The evolution of $\cos(\theta)$ for the same four orbits shown in Fig.~\ref{fig:torque}. Here, $\theta$ is the angle between the torque caused by the stripped particles and the angular momentum vector of the subhalo. The grey and pink shaded lines denote pericentric passages and apocentric passages, respectively. The angle oscillates over the course of each orbit as the geometry of the stripped material evolves with respect to the orbit of the subhalo. This oscillation appears connected to pericentric passages, likely because most of the stripped material resides near the center of the host halo.}
\label{fig:cosine}
\end{figure*}

The fact that $\rmd\ln j/\rmd t$ flips sign during pericentric passages, and occasionally also close to apocentric passage (this is most pronounced for the $\eta=0.3$ orbit shown in the top panel of Fig.~\ref{fig:torque}), indicates that the direction of the torque must vary with time. This is confirmed in Fig.~\ref{fig:cosine}, which plots, for the same orbits as in  Fig.~\ref{fig:torque}, the time evolution of the cosine of the angle, $\theta$, between the torque $\bmath{\calT}_{\rm SF}$ and the subhalo's orbital angular momentum $\bj$. If $\cos \theta < 0$, then $\calT$  opposes $\bj$, causing the subhalo to lose orbital angular momentum. In the opposite case, where $\cos \theta > 0$, the torque causes an increase in $j$. Furthermore, if $\vert\cos\theta\vert = 1$ the torque is either perfectly aligned or anti-aligned with $j$, causing only the amplitude of $\bj$ to change, but not its direction. If, on the other hand $\vert \cos\theta\vert < 1$, the misalignment between torque and angular momentum will cause a precession in the orbital plane of the subhalo, and thus a change in the direction of $\bj$. As is evident from  Fig.~\ref{fig:cosine}, in most cases, and during most of the time, the torque is purely retarding (i.e., $\cos\theta = -1$). However, there are also episodes when $\cos\theta$ fluctuates wildly. Occasionally, $\cos\theta$ will be equal to, or very similar to, $+1$ for a prolonged period; the most pronounced case in Fig.~\ref{fig:cosine} corresponds to a period around the first apocentric passage of the $\eta=0.3$ orbit (upper panel). During these periods the torque enhances the orbital angular momentum of the subhalo.

The drastic fluctuations in the angle between $\bmath{\calT}_{\rm SF}$ and $\bj$ are not entirely unexpected. After all, the torque arises from the stripped material, which can have a very complex morphology, and which evolves with time due to phase-mixing and due to the continued stripping of the subhalo. This aspect of self-friction not only distinguishes it from standard dynamical friction, which always acts in the direction opposite of the subhalo's velocity vector, corresponding to a retarding torque with $\cos\theta = -1$ (but see Just \& Penarrubia 2005), it also implies that an accurate analytical treatment of self-friction is likely to remain intractable. 

\subsection{Comparison of self-friction to traditional dynamical friction}
\label{sec:comp}

In this section, we aim to quantify the overall importance of self-friction, in particular as compared to ``traditional'' dynamical friction caused by (the constituent particles of) the host halo. We do so using a two-pronged approach; first, we compare the self-friction merging timescales derived above to dynamical friction merging timescales taken from the literature. Next, we compare the instantaneous torque of self-friction to the retarding torque expected from the \citet{Chandrasekhar1943} formula for dynamical friction from the host halo.

\subsubsection{Comparison of merging timescales}
\label{sec:comp_timescale}

\begin{figure}
\centering
\includegraphics[width = \columnwidth]{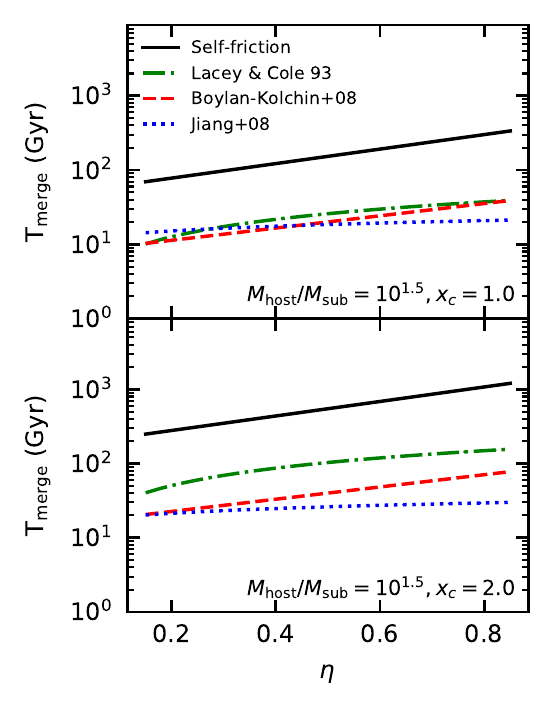}
\caption{An analytical comparison between the self-friction timescale fitting function and merging timescales found in the literature. We find that the self-friction merging time has similar scaling with $\eta$ and $M_{\rm host}/M_{\rm sub}$, but is consistently ${\sim}10$ times longer than the merging time for subhaloes from the literature. The \citet{Lacey1993} analytical description only includes dynamical friction due to the host halo and is multiplied by three to account for mass loss as suggested by \citet{Mo2010}. It is worth nothing that \citet{Boylan-Kolchin2008} and \citet{Jiang2008a} simulate both the subhalo and host halo and therefore naturally include the effects of both self-friction and dynamical friction.}
\label{fig:tau_comp}
\end{figure}

We compare the self-friction decay timescale derived in \S\ref{sec:ts} above to dynamical friction merging timescales taken from three different studies in the literature: \citet{Boylan-Kolchin2008}, who used idealized simulations of individual subhaloes orbiting a host halo, \citet{Jiang2008a} who derived merging times of subhaloes from a cosmological $N$-body simulation, and \citet{Lacey1993} who estimated merging times using an analytical model. Since the latter assumed a constant mass for the subhalo, we multiply their timescales by a factor of three to account for the impact of tidal mass loss \citep[see][section 12.3.1]{Mo2010}. We emphasize that the timescales of  \citet{Boylan-Kolchin2008} and \citet{Jiang2008a}  include the effects of both dynamical friction from the host halo, as well as self-friction (i.e., self-friction is always present in numerical simulations). The self-friction merging timescale used here is defined as an e-folding time under the assumption of an exponential decay of the subhalo's specific, orbital angular momentum, while the other timescales are all based on some estimate of the time it takes to `fully' merge (typically defined as the time it takes for either $j \rightarrow 0$ or $r \rightarrow 0$). In order to roughly account for these different definitions, we define the self-friction merging time as $T_{\rm merge, SF} = 2 \times \tau_{\rm SF}$, which corresponds to the time it takes the specific, orbital angular momentum, $j$, to decay (exponentially) to ${\sim}10\%$ of its original value. Using a definition $T_{\rm merge, SF} = 3 \times \tau_{\rm SF}$ instead does not significantly alter any of our main conclusions.

Fig.~\ref{fig:tau_comp} plots the various merging timescales as a function of $\eta$ for two different values of $x_\rmc$. The three merging timescales from the literature agree within a factor of ${\sim}2$ over the range of orbital parameters shown. In general, the behavior of the self-friction merging time scales similarly to studies that include dynamical friction due to the host except that it is roughly a factor of ten longer. As noted above the only major difference in scaling is the much stronger dependence on $x_\rmc$. Taken at face value, this suggests that the relative importance of self-friction is larger for more bound orbits. Since friction makes orbits more bound, it thus also suggests that self-friction becomes more important, in a relative sense, at the late stages of the orbital evolution of a subhalo.  However, we caution that our simulations of self-friction only cover the range $1.0 \leq x_\rmc \leq 2.0$, and that an extrapolation of $\tau_{\rm SF}(x_\rmc,\eta)$ outside of this range is to be taken with a grain of salt. 

\subsubsection{Comparison of instantaneous torques}
\label{sec:comp_model}

As another way to gauge the relative importance of self-friction, we compare the self-friction torque, $\calT_{\rm SF}$, to the torque responsible for traditional dynamical friction, $\calT_{\rm DF}$. We compute the latter using the well-known \citet{Chandrasekhar1943} formula for the dynamical friction deceleration 
\begin{equation}\label{eqn:CSDF}
\bmath{a}_{\rm DF} = -\frac{4\pi\, \ln\Lambda\, G^2 M_\rms}{v_\rms^2}\ \rho(R) \left[{\rm erf}(\rmX) - \frac{2 \rmX}{\sqrt{\pi}} e^{-\rmX^2} \right] \, \frac{\bmath{v}_\rms}{v_\rms},
\end{equation}
Here $M_\rms = M_{\rm sub}(t) = f_\rmb(t) M_{\rm sub}$ is the bound mass of the subhalo at time $t$, $v_\rms = V_{\rm COM}(t)$ is the COM velocity of the subhalo with respect to the host halo at time $t$, $R = R_{\rm COM}(t)$ is the orbital radius of the subhalo at time $t$, and $X = v_\rms(t)/V_{\rm circ}(R)$ with $V_{\rm circ}(R)$ the circular velocity of the host halo at halocentric radius $R$. The factor in square brackets expresses the fraction of host halo particles with a speed $v < v_\rms$, which, as is standard, has been computed under the assumption that the host particles follow a (locally) Maxwellian velocity distribution. For the Coulomb logarithm, given the uncertainties involved, we perform our computations using two different forms:
$\ln\Lambda = \ln[1 + M_{\rm host}/M_{\rm sub}(t)]$, which is the form adopted by \citet{Colpi1999}, \citet{Boylan-Kolchin2008}, \citet{Mo2010} and many others, and $\ln\Lambda =\ln[R(t)/r_\rmh(t)]$, with $r_\rmh(t)$ the half-mass radius of the bound remnant. The latter form is motivated by discussions in \citet{Hashimoto2003} and \citet{Petts2015}. Note that both forms for the Coulomb logarithm are time-dependent. Using our simulations, we compute $\bmath{a}_{\rm DF}$ as function of time using the instantaneous values of $M_\rms$, $v_\rms$, $R$, and $r_\rmh$ from which we then calculate the Chandrasekhar retarding torque $\calT_{\rm DF} =\bmath{R}_{\rm COM} \times \bmath{a}_{\rm DF}$.
\begin{figure*}
\centering
\includegraphics[width = \textwidth]{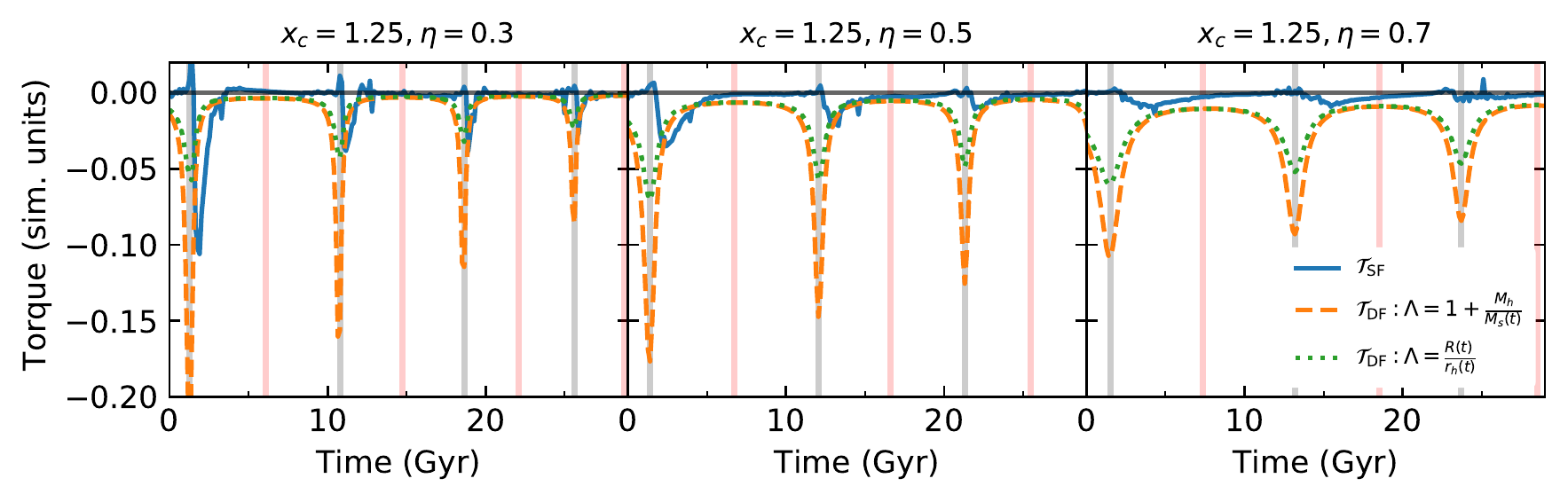}
\caption{The time evolution of the self-friction torque caused by the stripped particles, $\calT_{\rm SF}$ (blue solid curves), compared to the the evolution of the dynamical friction torque expected from the host, $\calT_{\rm DF}$. The latter is computed using the Chandrasekhar equation~\eqref{eqn:CSDF} for two assumed forms of the Coulomb logarithm (as indicated). The grey and pink shaded lines denote pericentric and apocentric passages, respectively. Although the self-friction torque can occasionally be larger than that due to dynamical friction from the host, integrated over an entire orbit it tends to be sub-dominant.}
\label{fig:inst_comp}
\end{figure*}

In Fig.~\ref{fig:inst_comp}, we compare the self-friction torque, projected along the direction of the specific, orbital angular momentum of the subhalo (blue, solid curves), to the dynamical friction torques $\calT_{\rm DF}$ obtained using $\ln\Lambda = \ln[M_{1 + \rm host}/M_{\rm sub}(t)]$ (orange, dashed curves) and $\ln\Lambda = \ln[R(t)/r_\rmh(t)]$ (green, dotted curves)\footnote{Since $\bmath{a}_{\rm DF}$ is always pointing in the direction opposite to $\bmath{v}_\rms$, these torque are always perfectly anti-aligned with $\bj$.}. Different panels correspond to different orbital parameters, as indicated, and all cases shown have $\mhs = 10$. All three torques reveal a periodic behavior that aligns with the orbital phase. In particular, the predicted torque due to dynamical friction from the host is strongest during pericentric passage, when the density of the host halo is largest and the subhalo is moving fastest. Note that the uncertainty in the Coulomb logarithm introduces an uncertainty in the torque of roughly a factor of two, which is most pronounced during pericentric passages. Interestingly, the self-friction torque is strongest shortly after the subhalo has passed its pericentre, which is proceeded by a short `burst' of positive (enhancing) torque right around pericentric passage. This rapid evolution in the self-friction torque during pericentric passage is to be expected from the fact that this is also the orbital phase during which most of the tidal mass stripping occurs. As the subhalo passes pericenter, the mass ratio and relative orientation of remnant and stripped material, and thus the torque, evolves rapidly.

A comparison of the self-friction torque with the expected dynamical friction torque shows once again that self-friction is clearly sub-dominant, at least when integrated over an entire orbit. In some cases (i.e., the more eccentric orbits), self-friction can briefly dominate over dynamical friction shortly after pericentric passage. Overall, though, it is clear that self-friction only contributes about 1-10 percent of the total friction experienced by the subhalo, and thus that one does not make a significant error when ignoring self-friction all together. 

\section{Summary and Discussion}
\label{sec:disc}

When an extended subject mass orbits a host system within which it experiences dynamical friction, it typically also experiences mass loss due to the tidal forces from the host. This tidally stripped material exerts a gravitational force on the bound remnant, which is typically retarding, thus giving rise to an additional friction force that we call self-friction. In order to investigate the relative importance of this self-friction compared to standard dynamical friction due to the host system, we have run a suite of idealized simulations of an $N$-body subhalo orbiting the static, analytical potential of a host halo. This has the advantage that it isolates the effect of self-friction. Our simulations cover the full range of orbital parameters relevant for dark matter substructure \citep{Jiang2015} and have mass ratios between host and subhalo at accretion (i.e., prior to any mass loss) that cover the range from 10 to 100. 

Without exception, the subhaloes in our simulations lose orbital angular momentum, which causes them to spiral towards the centre of the host halo. We have explicitly demonstrated that this is due to the torque exerted by the stripped material on the remaining bound subhalo material. Unlike Chandrasekhar's dynamical friction, which always acts in the direction opposite of the velocity vector of the subhalo, the torque due to self-friction can be misaligned with the orbital angular momentum vector, causing orbital precession. In fact, due to the complicated, time-dependent geometry of the stripped, phase-mixed material, the torque vector can occasionally align with the angular momentum vector, causing an accelerating, rather than decelerating, force. However, most of the time the self-friction torque is retarding, causing a net loss of orbital angular momentum.

We have quantified how the characteristic self-friction time, $\tau_{\rm SF}$, defined as the exponential decay time of the subhalo's specific orbital angular momentum, depends on the orbital energy (expressed in terms of $x_\rmc$), orbital angular momentum (expressed in terms of the circularity $\eta$) and initial mass ratio $M_{\rm host}/M_{\rm sub}$. Typically, self-friction is more pronounced for more bound orbits ($\tau_{\rm SF} \propto x_\rmc^{1.9}$), for more radial orbits ($\tau_{\rm SF} \propto \exp[2.3\eta]$), and for more massive subhaloes ($\tau_{\rm SF} \propto [M_{\rm host}/M_{\rm sub}]^{1.2}$). Except for the dependence on $x_\rmc$, which is stronger in the case of self-friction, these scalings are similar to those of the standard dynamical friction time measured in simulations with a live host halo \citep[cf.,][]{Boylan-Kolchin2008}. However, the latter is typically about an order of magnitude shorter than the self-friction time, indicating that self-friction contributes roughly at the 10 percent level, in reasonable agreement with a previous estimate by \citet{Fujii2006}.

For a given initial mass ratio, we find that the fraction of orbital angular momentum lost up to some time $t$ is tightly correlated with the amount of mass loss that the subhalo has experienced hitherto, independent of the orbital parameters. On the other hand, at fixed host halo mass and fixed bound mass fraction, subhaloes that are more massive have lost a larger fraction of their initial, orbital angular momentum due to self-friction. Those subhaloes, though, will also lose a larger fraction of their initial orbital angular momentum due to dynamical friction, such that the fractional contribution of self-friction remains at about the 10 percent level.


\section*{Acknowledgements}

The authors thank Uddipan Banik, Dhruba Dutta Chowdhury, Johannes Lange, Nir Mandelker and Michael Tremmel for useful discussion. TBM thanks the Gruber Foundation and Patricia Gruber for their generous support of the work presented here. FvdB is supported by the National Aeronautics and Space Administration through Grant Nos. 17-ATP17-0028 and 19-ATP19-0059 issued as part of the Astrophysics Theory Program, and received addition support from the Klaus Tschira foundation. SBG is supported by the US National Science Foundation Graduate Research Fellowship under Grant No. DGE-1752134. GO acknowledges funding from the European Research Council (ERC) under the European Union's Horizon 2020 research and innovation programme (grant agreement no. 679145, project `COSMO-SIMS').


\bibliographystyle{mnras}
\bibliography{self_fric} 


\appendix

\section{Numerical reliability of simulation results}
\label{app:num}
\begin{figure*}
\centering
\includegraphics[width = 0.99\textwidth]{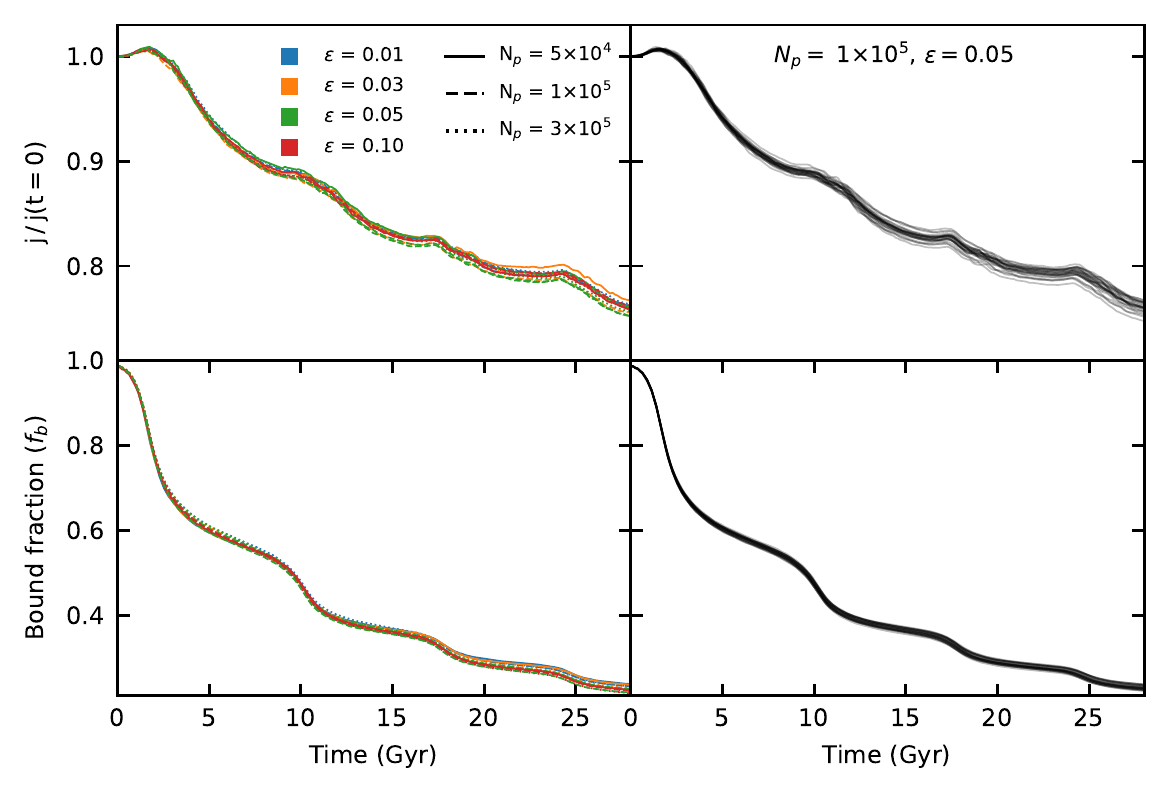}
\caption{\textit{Top:} The evolution of the specific angular momentum, $j$, for two sets of numerical tests. In the first set (\textit{left}), we use the same initial conditions but vary the number of particles in the subhalo and the softening length, $\epsilon$. In the second set (\textit{right}), we keep the numerical parameters fixed to their fiducial values but vary the random seed used to generate the subhalo initial conditions 25 times. \textit{Bottom:} The same two sets of numerical tests, now plotting the evolution of the bound fraction, $f_\rmb$, over the course of the orbital evolution. We find very little difference (less then 5\% variation) between any of the runs, illustrating that our results are numerically stable.}
\label{fig:num}
\end{figure*}

In order to test and assure that our simulation results are robust and reliable, we have performed a number of tests. The left-hand panels of Fig.~\ref{fig:num} show the time evolution of the orbital angular momentum (upper panels) and the bound mass fraction (lower panels) for a set of simulations that only differ in the value of the softening length, $\epsilon$, and the number of particles, $N_\rmp$, as indicated. All other parameter are kept fixed to the fiducial values used for the simulation shown in Fig.~\ref{fig:intro}. As shown in \citetalias{VandenBosch2018a} a softening length that is too large can result in too much mass loss, which ultimately leads to artificial disruption of the subhalo, while inadequate mass resolution (i.e., too few particles) can result in in a runaway instability triggered by the amplification of discreteness noise in the presence of a tidal field. As is evident, though, our simulation results are robust to changes in $\epsilon$ or $N_\rmp$, and are thus free from such numerical artifacts. 

One of the manifestations of discreteness noise highlighted in \citetalias{VandenBosch2018a} is the fact that different random realizations of the same simulation (i.e., simulations that only differ in the random seed used to set up the initial phase-space coordinates for the subhalo particles), can result in very different time-evolutions of the bound-mass fraction. In order to test for such discreteness noise issues in our simulations, we have run 25 simulations of the same set-up (orbital parameters, softening length, number of particles, etc) as used for our fiducial simulation shown in Fig.~\ref{fig:intro}. The corresponding time evolution of the orbital angular momentum and the bound mass fraction are shown in the right-hand panels of Fig.~\ref{fig:num}. Again, we find the evolution of $j$ and $f_\rmb$ to be stable with less then 5\% variation among different simulations after 30 Gyr.

\bsp	
\label{lastpage}

\end{document}